\documentclass[twocolumn,showpacs,preprintnumbers,amsmath,amssymb]{revtex4}

\usepackage{graphicx}
\usepackage{dcolumn}
\usepackage{bm}
\usepackage{natbib}

\begin{document}


\title{
The origin of the phase separation in partially deuterated $\kappa$-(ET)$_2$Cu[N(CN)$_2$]Br studied by infrared magneto-optical imaging spectroscopy
}

\author{T. Nishi$^{1}$}\thanks{E-mail: tnishi@ims.ac.jp}
\author{S. Kimura$^{1,2}$}
\author{T. Takahashi$^{3}$}
\author{T. Ito$^{2}$}
\author{H.J. Im$^{1}$}
\author{Y.S. Kwon$^{2,4}$}
\author{K. Miyagawa$^{5}$}
\author{H. Taniguchi$^{6}$}
\author{A. Kawamoto$^{7}$}
\author{K. Kanoda$^{5}$}

\affiliation{
$^1$Department of Structural Molecular Science, The Graduate University for Advanced Studies, Okazaki 444-8585, Japan\\
$^2$UVSOR Facility, Institute for Molecular Science, Okazaki 444-8585, Japan\\
$^3$Research Reactor Institute, Kyoto University, Osaka 590-0494, Japan\\
$^4$Department of Physics, SungKyunKwan University, Suwon 440-746, South Korea\\
$^5$Department of Applied Physics, The University of Tokyo, Tokyo 113-8656, Japan\\ and CREST, Japan Science and Technology Corporation, Japan\\
$^6$Department of Physics, Saitama University, Saitama 338-8570, Japan\\
$^7$Department of Physics, Hokkaido University, Sapporo 060-0810, Japan
}

\date{\today}

\begin{abstract}
The direct observation of the phase separation between the metallic and insulating states of 75~\%-deuterated $\kappa$-(ET)$_2$Cu[N(CN)$_2$]Br ($d33$) using infrared magneto-optical imaging spectroscopy is reported, as well as the associated temperature, cooling rate, and magnetic field dependencies of the separation.
The distribution of the center of spectral weight ($\langle~\omega~\rangle$) of $d33$ did not change under any of the conditions in which data were taken and was wider than that of the non-deuterated material.
This result indicates that the inhomogenity of the sample itself is important as part of the origin of the metal - insulator phase separation.
\end{abstract}

\pacs{71.27.+a, 78.20.Ls, 75.30.Kz}
\maketitle

Since organic conductors have a variety of physical properties, many research works have been performed for a long time.
In the organic conductors, quasi-two-dimensional materials attract attention for their similar physical properties to high-$T_{\rm C}$ cuprates.
The ground state of the quasi-two-dimensional organic conductor, $\kappa$-(ET)$_2X$ (ET = bis(ethylenedithio)-tetrathiafulvalene, $X$ = Cu[N(CN)]$_2$Br, Cu[N(CN)]$_2$Cl, Cu(NCS)$_2$ and so on) in this case, is determined by the universal parameter $U/W$, where $U$ and $W$ are the on-site Coulomb energy and band width, respectively, of the upper HOMO band of, in this case, the ET dimer.~\cite{ref01}
This type of material is classified as a Mott insulator system.  In the $\kappa$-(ET)$_2X$, especially $X$ = Cu[N(CN)$_2$]Br, the superconducting (S) state directly changes to the antiferromagnetic (AFI) state below about 11~K with increasing $U/W$ at around $U/W = 1$.~\cite{ref02, ref03, ref04}
The $U/W$ parameter can be controlled by the deuteration and cooling rate.~\cite{ref05}
At the boundary, it is believed that the phase separation with $\mu$m-scale domains between S and AFI states appears.
But there is no experimental result that has directly verified this observation.~\cite{ref06}
In this paper, we report the direct observation of the phase separation and its temperature, cooling rate and magnetic field dependences.

Partially deuterated $\kappa$-(ET)$_2$Cu[N(CN)$_2$]Br shows characteristic physical properties.
In 75~\% deuterated material (denoted as $d33$ hereafter), the phase changes from S to AFI when the cooling rate is increased, and also a magnetically induced S - AFI transition is observed in 50~\%-deuterated material.~\cite{ref06, ref07, ref08}
Both of these transitions are considered to relate to the phase separation.~\cite{ref09}
Especially, $d33$ indicates the partial S state in the fast cooling condition observed in the ac susceptibility experiment.
Both the size of the separated phases and the electronic behavior of the crossover between the S and AFI phases are unsettled questions.~\cite{ref06}
For example, in high-$T_{\rm C}$ cuprates, the phase separation on the nanometer scale has been observed by scanning tunneling microscopy.~\cite{ref10}
Contrary to this, the domain size of $\kappa$-(ET)$_2X$ is believed to be on the $\mu$m-scale, although again there is no information regarding the specifics of this.

To investigate metal - insulator domains like these on the $\mu$m-scale and the temperature and magnetic field dependencies of the associated transitions, Kimura {\it et al.} constructed an infrared magneto-optical imaging spectroscopy apparatus.~\cite{ref11, ref12}
The main purpose of this apparatus is to investigate the spatial distribution of optical reflectivity spectra as well as the electronic structure near the Fermi level at low temperatures and magnetic fields.
Using this apparatus, infrared reflectivity spectra in the presence of magnetic fields and their spatial contribution can be obtained, and the spatial distribution of metallic and insulating states can be observed based on the spectral shape.
Based on these methods, we firstly observed here the phase separation of the metallic and insulating states in $d33$ in contrast to no phase separation in the non-deuterated material ($d00$), and the temperature, cooling rate, and magnetic field dependencies of those transitions.

Single crystals of partially- and non-deuterated $\kappa$-($dnn$-ET)$_2$Cu[N(CN)$_2$]Br were prepared by a conventional electrochemical oxidation method.
The samples were placed on a substrate using a silver paste that avoided putting stress on the samples.  The optical reflectivity measurements between $\omega$ = 800 - 8000~cm$^{-1}$ were performed at the infrared magneto-optical station of the beam line 43IR of a synchrotron radiation facility, SPring-8, in Hyogo, Japan.~\cite{ref11, ref12}
The two cooling rates for both of $d00$ and $d33$ were set to 17~K/min for fast cooling (fc), and 0.05~K/min for slow cooling (sc) in the temperature range of 90 - 70~K.
Two-dimensional imaging data were obtained by measuring each sample with 12~$\mu$m steps in the same area with a spatial resolution of less than 20~$\mu$m.
All spectra were recorded using unpolarized light because the overall spectral change due to the outer perturbations is discussed, though polarized reflectivity spectra are usually used for the analysis of the electronic and phonon structures.

\begin{figure}[t]
\begin{center}
\includegraphics[width=8cm]{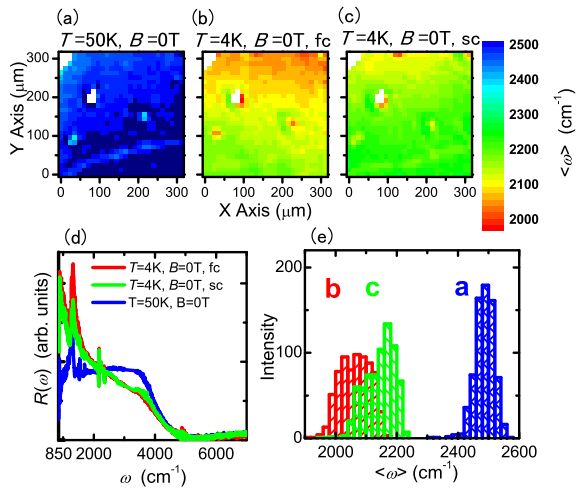}
\end{center}
\caption{
(Color) Spatial distribution of the center of spectral weight ($\langle~\omega~\rangle$) of $d00$ at 50~K, 0~T (a), 4~K, 0~T in the fast cooling condition (b) and 4~K, 0~T in the slow cooling condition (c).
(d) Reflectivity spectra at the same point of $d00$ near the center of the image under the conditions in (a)-(c).
Metallic and insulating reflectivity spectra appear at 4 and 50~K at ambient magnetic field, respectively.
(e) Statistic distribution of $\langle~\omega~\rangle$ for the conditions in (a)-(c).
}
\label{d00}
\end{figure}
The results for $d00$ at $T =$ 50 and 4~K at $B =$ 0~T are shown in Figure~\ref{d00}.
This material is in the S and paramagnetic insulating (PI) phases at 4 and 50~K, respectively.
The physical characteristics are reflected in the temperature dependence of the infrared reflectivity spectrum in Fig.~\ref{d00}~(d).
The spectra at 4~K, 0~T in both cooling conditions have metallic reflectivities in the low wave number region and that at 50~K, 0~T has an insulating reflectivity based on its almost constant reflectivity below 2000~cm$^{-1}$ except for the TO phonons at around 1300~cm$^{-1}$.
These spectral shapes are consistent with previous work.~\cite{ref13, ref14}
Because of the conservation of spectral weight, the reflectivity at around 3000~cm$^{-1}$ drops and rises at 4 and 50~K, respectively.
This behavior corresponds to the temperature dependence of the electronic structure of the lower- and upper-Hubbard bands as described by a typical Mott transition.
To clarify the change in the reflectivity spectrum as well as the change in the electronic structure, the center of spectral weight of the reflectivity spectrum ($\langle~\omega~\rangle$) was obtained by using the following function:
\[
\langle~\omega~\rangle = \frac{\int_{0}^{\omega_2} \omega R(\omega) d\omega}{\int_{0}^{\omega_2} R(\omega) d\omega}.
\]
Here, the integration area was limited below $\omega_2$ = 5000~cm$^{-1}$ in which the sum rule is satisfied.
In an ordinary case, the sum rule is evaluated from optical conductivity spectra derived from the Kramers-Kronig analysis of the reflectivity spectra.
However, to clarify the spectral change due to a difference in the conditions, we employed the reflectivity spectra to determine $\langle~\omega~\rangle$.

The spatial distributions of $\langle~\omega~\rangle$ at 50~K, 0~T, at 4~K, 0~T in the fc condition and at 4~K, 0~T in the sc condition are shown in Figs.~\ref{d00}~(a), (b) and (c), respectively.
In comparison between Figs.~\ref{d00}~(b) and (c), the cooling rate effect is small in $d00$.
Judging from the physical properties at 4 and 50~K, the blue-colored region is the insulating phase and the colors from green to red are the metallic phase.
At 50~K, the material is clearly in a uniform insulating phase.
At 4~K, it must change to a uniform S phase in the both cooling condition.
However, the color is not uniform.
This indicates that the electronic structure on the sample surface is not constant.
Note that since the reflectivity spectrum in the S phase is a metallic one, the S and metallic states cannot be distinguished by this method.
The difference between the metallic and insulating states, however, can be clarified.

Figure~\ref{d00}~(e) indicates the $\langle~\omega~\rangle$ distribution in the areas indicated in Figs.~\ref{d00}~(a), (b) and (c).
The center and full width ($2\sigma$) of $\langle~\omega~\rangle$ are 2486 and 60~cm$^{-1}$ at 50~K, 0~T, 2066 and 94~cm$^{-1}$ at 4~K, 0~T in the fc condition and 2150 and 97~cm$^{-1}$ at 4~K, 0~T in the sc condition, respectively.
The $\langle~\omega~\rangle$ distributions at 4~K, 0~T in both conditions are larger than that at 50~K, 0~T.
This means that the electronic structure becomes inhomogenity with decreasing temperature.
The $\langle~\omega~\rangle$ distributions at 50 and 4~K are clearly separated from each other.
Based on this, the phase transition from insulator to metal (superconductor) must occur over the whole sample surface of $d00$.
Figure~\ref{d00}~(d) indicates that the $\langle~\omega~\rangle$ distribution is completely separated at the boundary of 2300~cm$^{-1}$.
For future reference, the boundary between the metallic and insulating states is hereafter referred to as $\omega_{MI}$.
Note that the width of the $\langle~\omega~\rangle$ distribution indicates the inhomogenity in each phase and can be used for comparison with that of $d33$.

\begin{figure}[t]
\begin{center}
\includegraphics[width=8cm]{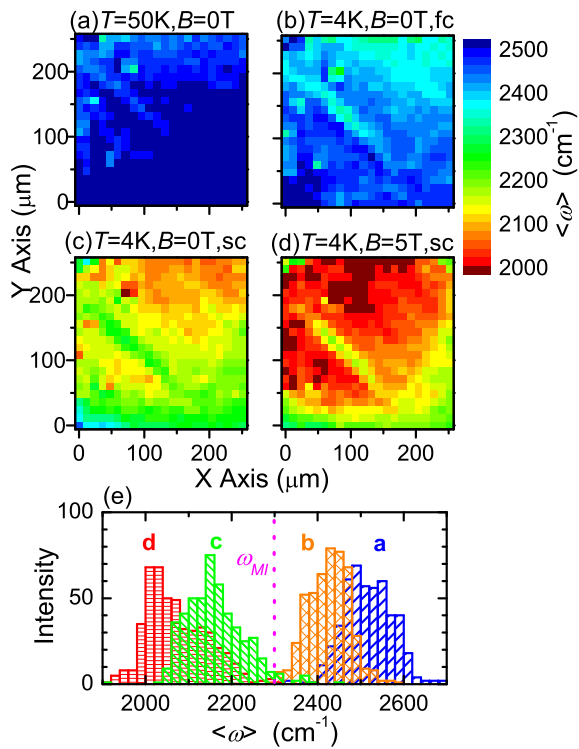}
\end{center}
\caption{
(Color)
Spatial distributions of $\langle~\omega~\rangle$ of $d33$ at 50~K, 0~T (a), 4~K, 0~T in the fast cooling  condition (b), 4~K, 0~T in the slow cooling condition (c) and 4~K, 5~T in the slow cooling condition (d).
There were no difference at 50~K, 0~T in (a) between the two cooling rates.
(e) Statistic distributions of $\langle~\omega~\rangle$ for conditions in (a) - (d).
$\omega_{MI}$ is the metal - insulator boundary evaluated from $d00$.
(See text for detail.)
}
\label{d33}
\end{figure}
The same analysis was performed on $d33$ and the results are shown in Figure~\ref{d33}.
The material is in the S, AFI, PI and paramagnetic metallic (PM) phases at 4~K, 0~T in the sc condition, at 4~K, 0~T in the fc condition, at 50~K, 0~T and at 4~K, 5~T in the sc condition, respectively.
The magnetic field of 5~T that is above $H_{c2}$ ($\sim$ 4~T) at 4~K was applied perpendicular to the $ac$-plane.
Figure~\ref{d33}~(a) shows the result at 50 K in the sc condition.
There was no difference between the sc and fc cases.
The results indicate there was no difference between the two cooling rates.
This means that the PI state is not affected by the cooling rate.
The whole area is in the insulating phase based on $\langle~\omega~\rangle$ for this phase being distinctly higher than $\omega_{MI}$, as shown in Figure~\ref{d33}~(e).
The center wave number of $\langle~\omega~\rangle$ at 50~K, 0~T is about 2510~cm$^{-1}$, similar to that observed for $d00$.
Therefore, the electronic structure of the PI phase in $d33$ are the same as those in $d00$.
However, the $2\sigma$ of the $\langle~\omega~\rangle$ distribution is about 120~cm$^{-1}$, much larger than that for $d00$.
Assuming the width indicates the level of inhomogenity, the sample surface of $d33$ is clearly more inhomogeneous than that of $d00$.

In the fc condition, $d33$ is nearly in the insulating phase at 4~K, 0~T as shown in Fig.~\ref{d33}~(b).
This is consistent with the observation that $d33$ is in the AFI phase under these conditions.
However, the center of $\langle~\omega~\rangle$ is slightly shifted to the metallic (lower wave number) side, as shown in Fig.~\ref{d33}~(e).
This puts the AFI phase closer to the $\omega_{MI}$ boundary, which is consistent with the location of the AFI phase being in the vicinity of the Mott boundary.
The width of the $\langle~\omega~\rangle$ distribution is similar to that at 50~K, 0~T in Fig.~\ref{d33}~(a).
The tail of the $\langle~\omega~\rangle$ distribution crosses $\omega_{MI}$.
This indicates that a tiny area in the sample surface changes to metallic state, {\it i.e.}, the phase separation between metal and insulator appears.
The metallic phase can be evaluated to be 2.5~\% based on the percentage of $\langle~\omega~\rangle$ below $\omega_{MI}$.

As shown in Fig.~\ref{d33}~(c), the phase in the sc condition at 4~K, 0~T is almost metallic based on the $\langle~\omega~\rangle$ distribution shifting to around 2150~cm$^{-1}$ below $\omega_{MI}$.
The center wave number is equal to that of $d00$ in the sc condition.
This means that the electronic structure of $d33$ in the S state is the same as that of $d00$.
However, the width is larger than that of $d00$.
The width is similar to those at 50~K, 0~T and also at 4~K, 0~T in the fc condition.
This indicates similar inhomogenity under all conditions.
As shown in Fig.~\ref{d33}~(e), the tail of the high wave number side of the $\langle~\omega~\rangle$ distribution crosses $\omega_{MI}$.
The volume fraction of the metallic phase that is evaluated based on the percentage of the $\langle~\omega~\rangle$ below $\omega_{MI}$ is 99~\%.
In comparison with the fc condition in Fig.~\ref{d33}~(b), the metallic volume fraction in the sample increases with a decreasing cooling rate.
This is qualitatively consistent with the result of the ac susceptibility experiment by Taniguchi {\it et al.}~\cite{ref06}
However, if the metallic area is regarded as a perfectly superconducting phase, the value is much higher than that received by the ac susceptibility measurement.
To explain the controversy, the PI phase is considered to be present in part of the metallic area.

When a magnetic field is applied, the S state collapses and changes to a PM state as indicated in electrical resistivity data.~\cite{ref08, ref15}
As shown in Fig.~\ref{d33}~(d), the spatial image of $\langle~\omega~\rangle$ shifts to the red color side, which indicates the phase changed to very metallic.
$\langle~\omega~\rangle$ indicates the spectral shape as well as the electronic structure near the Fermi level including the lower- and upper-Hubbard bands.
This indicates that the electron correlation is reduced by a magnetic field.
As shown in Fig.~\ref{d33}~(e), the center of the $\langle~\omega~\rangle$ distribution shifts to the low wave number side, but the width is almost the same as that under other conditions.

\begin{figure}[t]
\begin{center}
\includegraphics[width=6.0cm]{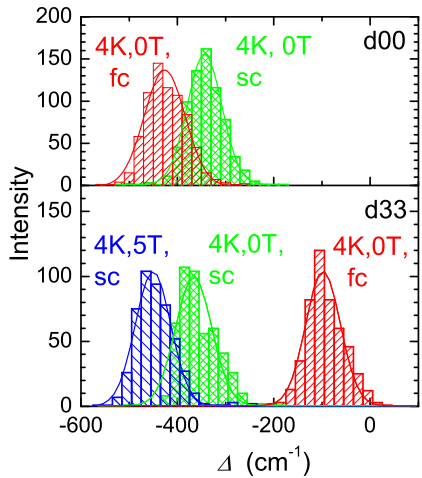}
\end{center}
\caption{
(Color)
Bars: Statistic distributions of the change in $\langle~\omega~\rangle$ in each condition from that in 50~K, 0~T at the same points.
The lines are Gaussian fitted curves.
$\Delta$ indicates $\langle~\omega~\rangle_{T,B}$ - $\langle~\omega~\rangle_{50K,0T}$.
}
\label{diff}
\end{figure}
To evaluate the different wave number in $\langle~\omega~\rangle$ due to each condition, the difference ($\Delta$) from that in 50~K, 0~T at the same measured points were derived from Figs.~\ref{d00} and \ref{d33}.
The statistic distribution is plotted in Fig.~\ref{diff}.
In the case of $d00$, the character in the fc condition is more metallic than that in the sc condition with a similar $\Delta$ shape (the $2\sigma$ is 84~$\pm$~6~cm$^{-1}$ for fc and 72~$\pm$~3~cm$^{-1}$ for sc).
Since the $\Delta$ width is smaller than the $\langle~\omega~\rangle$ one ($\sim$ 90~cm$^{-1}$) in Fig.~\ref{d00}, the phase changes uniformly.
On the other hand, in the case of $d33$, all of the widths of the $\Delta$ distribution are 73 $\pm$ 5~cm$^{-1}$, which is much lower than the $\langle~\omega~\rangle$ widths shown in Fig.~\ref{d33}~(e).
This means that the shift values for each point are almost the same, {\it i.e.}, the spatial inhomogenity shown in Figs.~\ref{d33} (a) - (d) originates from the inhomogenity contained in the material itself.
The width of $\Delta$ for $d33$ is similar to that of $d00$, and is almost constant for all conditions in $d33$ and also in $d00$.
This narrow $\langle~\omega~\rangle$ distribution for $d00$ indicates that the phase transition was highly uniform.
On the other hand, the wide $\langle~\omega~\rangle$ distribution for $d33$ explains the metal-insulator phase transition, {\it i.e.}, the most metallic (insulating) side remains in the metallic (insulating) phase even though the other area changes to the insulating (metallic) phase.
Therefore the inhomogenity contained in the partial deuterated material itself is concluded to be the origin of the metal (superconductor) - insulator phase separation.

Finally, the domain size of the metal and insulator states is discussed.
The experimental results indicate that the sample surface of $d33$ itself is more inhomogeneous than that of $d00$ because the width of the $\langle~\omega~\rangle$ distribution of $d33$ is larger than that of $d00$.
If the domain size is larger than the spatial resolution of about 20 $\mu$m in this experiment, the $\langle~\omega~\rangle$ distribution splits into two parts across $\omega_{MI}$.
However, since the $\langle~\omega~\rangle$ distribution of $d33$ is not a double peak structure but a broad single peak, the domain size is smaller than the experimental resolution of 20~$\mu$m.
However, the $\langle~\omega~\rangle$ distribution should be narrow in the case of a very small domain size because no contrast appears.
According to a simple simulation performed to explain the wide $\langle~\omega~\rangle$ distribution of $d33$, the domain size was evaluated to be about one order smaller than the spatial resolution, {\it i.e.}, the domain size must be a few $\mu$m.
To clarify the actual domain size, the normal infrared microscopy is not suitable because of the diffraction limit but a near-field infrared experiment with the sub-$\mu$m resolution may give us useful information.

To summarize, the direct observation of the phase separation of 75~\%-deuterated $\kappa$-(ET)$_2$Cu[N(CN)$_2$]Br ($d33$) was observed and compared with the non-deuterated material ($d00$) using infrared magneto-optical imaging spectroscopy.
The sample surface of $d33$ was found to be more inhomogeneous than that of $d00$.
This inhomogenity was consistent under different conditions, including under magnetic fields.
The phase separation in $d33$ originates from the inhomogenity contained in the material.
The domain size of the metallic and insulating states in $d33$ at the ground state is concluded to be a few $\mu$m.
The reflectivity spectrum of d33 at 5~T was more metallic than that at 0~T.
This reason is that the electron correlation plays an important role for the superconducting state at 0~T.

We would like to thank SPring-8 BL43IR staff members for their technical support.
The experiments were performed with the approval of the Japan Synchrotron Radiation Research Institute (Proposal Nos. 2002B0041-NS1-np and 2003A0076-NS1-np).


\end{document}